\documentclass[aps,prl,twocolumn,groupedaddress]{revtex4-1}
\usepackage{amsmath}
\usepackage{graphicx}
\DeclareGraphicsExtensions{.pdf,.eps,.png,.jpg,.mps}
\usepackage{epstopdf}
\usepackage{threeparttable}
\usepackage{color}
\begin{document}
\title{Broadband, electrically tuneable, third harmonic generation in graphene}
\author{G. Soavi$^1$, G. Wang$^1$, H. Rostami$^2$, D. Purdie$^1$, D. De Fazio$^1$, T. Ma$^1$, B. Luo$^1$, J. Wang$^1$, A. K. Ott$^1$, D. Yoon$^1$, S. Bourelle$^1$, J. E. Muench$^1$, I. Goykhman$^1$, S. Dal Conte$^{3,4}$, M. Celebrano$^{4}$, A. Tomadin$^2$, M. Polini$^2$, G. Cerullo$^{3,4}$, A. C. Ferrari$^1$}
\affiliation{$^1$ Cambridge Graphene Centre, University of Cambridge, Cambridge CB3 0FA, UK}
\affiliation{$^2$ Istituto Italiano di Tecnologia, Graphene Labs, Via Morego 30, I-16163 Genova, Italy}
\affiliation{$^3$ IFN-CNR, Piazza L. da Vinci 32, I-20133 Milano, Italy}
\affiliation{$^4$ Dipartimento di Fisica, Politecnico di Milano, Piazza L. da Vinci 32, I-20133 Milano, Italy}

\begin{abstract}
Optical harmonic generation occurs when high intensity light ($>10^{10}$W/m$^{2}$) interacts with a nonlinear material. Electrical control of the nonlinear optical response enables applications such as gate-tunable switches and frequency converters. Graphene displays exceptionally strong-light matter interaction and electrically and broadband tunable third order nonlinear susceptibility. Here we show that the third harmonic generation efficiency in graphene can be tuned by over two orders of magnitude by controlling the Fermi energy and the incident photon energy. This is due to logarithmic resonances in the imaginary part of the nonlinear conductivity arising from multi-photon transitions. Thanks to the linear dispersion of the massless Dirac fermions, ultrabroadband electrical tunability can be achieved, paving the way to electrically-tuneable broadband frequency converters for applications in optical communications and signal processing.
\end{abstract}

\maketitle
The response of a material to interaction with an optical field can be described by its polarization\cite{Shen1984}:
\begin{equation}
\vec{P} = \epsilon_0 [\chi^{(1)} \cdot \vec{E}+\chi^{(2)} : \vec{E}\vec{E}+\chi^{(3)}\vdots \vec{E}\vec{E}\vec{E}+ \cdots ]
\label{eq:Pol}
\end{equation}
where $\vec{E}$ is the incident electric field and $\epsilon_{0}$ is the permittivity of free space. $\chi^{(1)}$(dimensionless) is the linear susceptibility, while the tensors $\chi^{(2)}$ [m/V] and $\chi^{(3)}$ [m$^{2}$/V$^{2}$] are the second- and third-order nonlinear susceptibilities\cite{units}. Thanks to the nonlinear terms of $\vec{P}$, new frequencies can be generated inside a material due to harmonic generation\cite{FranPRL1961} and frequency mixing\cite{StolAPL1974}. E.g., in Second Harmonic Generation (SHG) an incident electromagnetic wave with angular frequency $\omega_{0}=2\pi\nu$, with $\nu$ the photon frequency, generates via $\chi^{(2)}$ a new electromagnetic wave with frequency $2\omega_{0}$\cite{FranPRL1961}. The SHG efficiency (SHGE) is defined as the ratio between the SH intensity and the intensity of the incoming light. Analogously, Third Harmonic Generation (THG) is the emission of a photon with energy triple that of the incident one. The THG efficiency (THGE) is defined as the ratio between the TH intensity and the intensity of the incoming light. Second-order nonlinear processes are also known as three-wave-mixing, as they mix two optical fields to produce a third one\cite{ArmsPhysRev1962}. Third-order nonlinear processes are known as four-wave-mixing (FWM)\cite{ArmsPhysRev1962}, as they mix three fields to produce a fourth one.

Nonlinear optical effects are exploited in a variety of applications, including laser technology\cite{SteinScience1999}, material processing\cite{ChanJLA1998} and telecommunications\cite{GarmOptExp2013}. E.g., to generate new photon frequencies (532nm from SHG in a Nd:YAG laser at 1.06$\mu$m)\cite{MillOptLett1997} or broadly tuneable ultrashort pulses (fs-ps) by optical parametric amplifiers (OPAs)\cite{CeruRSI2003} and optical parametric oscillators (OPOs)\cite{BoseOptLEtt1996}. High harmonic generation is also used for extreme UV light\cite{CorkPRL1993} and attosecond pulse generation\cite{CorkNP2007}, while difference frequency generation is used to create photons in the THz range\cite{FergNM2002}.

Second order nonlinear effects can only occur in materials without inversion symmetry, while third order ones occur in any system independent of symmetry\cite{Boyd2003}, and they thus represent the main intrinsic nonlinear response for most materials. THG intensity enhancement was achieved by exploiting magnetic dipole\cite{ShchNL2014} and excitonic resonances\cite{ChenPRB1993}, surface plasmons in Ag films\cite{TsanOptLett1996} and photonic-crystal waveguides\cite{CorcNP2009}, by exploiting spatial compression of the optical energy, resulting in an increase of the local optical field. Nonlinear optical effects depend on the characteristics of the impinging light beam(s) (frequency, polarization) and on the properties of the nonlinear material, dictated by its electronic structure. The ability to electrically control the nonlinear optical response of a material by a gate voltage opens up disruptive applications to compact nanophotonic devices with novel functionalities. However, to the best of our knowledge, external electrical control of the THGE has not been reported to date in any material.
\begin{figure}
\centerline{\includegraphics[width=90mm]{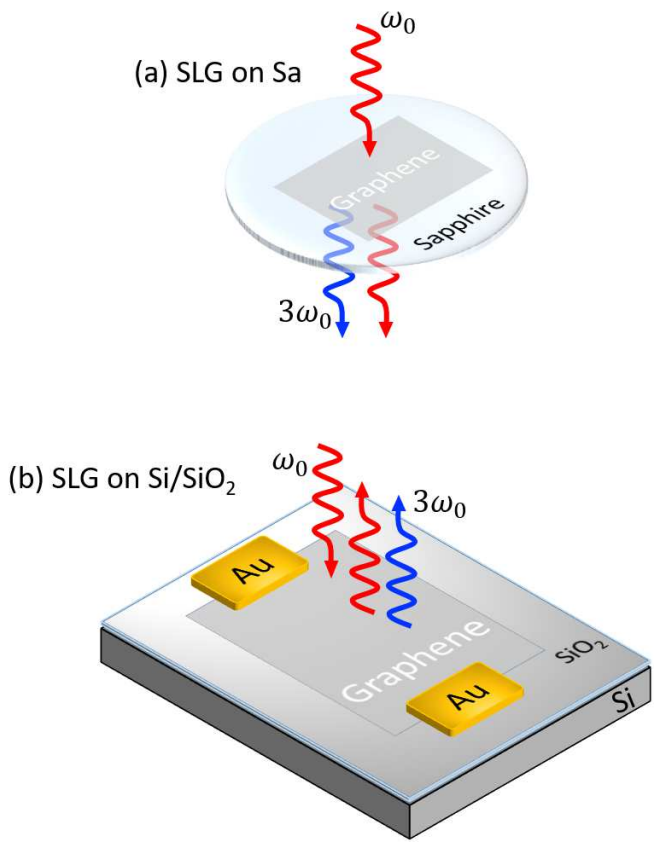}}
\caption{\textbf{Samples used for THG experiments.} (a) CVD SLG on Sa for transmission and (b) exfoliated SLG on Si/SiO$_{2}$ for reflection measurements.}
\label{fig:samples}
\end{figure}

Layered materials (LMs) have a strong nonlinear optical response\cite{SeylNatNano2015,HendPRL2010,WangPRL2015,LiuNatPhys2017,SaynNC2017,KleinNanoLett2017,SunACSNano2010,BonaNP2010}. Electrically tunable SHG was reported for monolayer WSe$_{2}$ for photon energies close to the A exciton ($\sim$1.66eV)\cite{SeylNatNano2015}. However, the tunability was limited to a narrow band ($\sim$10meV) in the proximity of the excitonic transition. Electrically tunable SHG was also reported by inversion symmetry breaking in bilayer MoS$_{2}$ close to the C exciton ($\sim$2.75eV)\cite{KleinNanoLett2017}, but SHGE was strongly dependent on the laser detuning with respect to the C exciton transition energy. Thus, in both cases electrical control was limited to narrow energy bands ($\sim$10-100meV) around the excitonic transitions. Graphene, instead, can provide electrically tunable nonlinearities over a much broader bandwidth thanks to the linear dispersion of the Dirac Fermions. In single layer graphene (SLG), SHG is forbidden due to symmetry\cite{MikhPRB2011,DeanPRB2010,AnPRB2014,AnNanoLett2013}. SHG was reported in the presence of an electric current\cite{AnNanoLett2013,AnPRB2014}, but weak compared to third-order nonlinear effects. The data in Refs.\citenum{HendPRL2010},\citenum{AnPRB2014} imply that THG in SLG is at least one order of magnitude stronger than SHG activated by inversion symmetry breaking. Thus, third order nonlinear effects are the most intense terms of the intrinsic nonlinear optical response of SLG.

Third order nonlinearities in SLG were studied both theoretically\cite{ChengNJPhys2014,ChengPRB2015,MikhPRB2016,RostPRB2016,RostPRB2017} and experimentally\cite{HendPRL2010,KumaPRB2013,AlexCondMat2017,HongPRX2013}. Ref.\citenum{HendPRL2010} reported that SLG has $\chi^{(3)}\sim$10$^{-15}$m$^{2}$/V$^{2}$ ($\sim$10$^{-7}$esu), several orders of magnitude higher than typical metals (\emph{e.g.} $\chi^{(3)}\sim 7.6 \times 10^{-19}$m$^{2}$/V$^{2}$ in Au\cite{Boyd2003}) and dielectrics (\emph{e.g.} $\chi^{(3)}\sim 2.5 \times 10^{-22}$m$^{2}$/V$^{2}$ in fused silica\cite{Boyd2003}). Ref.\citenum{HendPRL2010} also reported a $1/\omega_{0}^{4}$ proportionality of the third order optical nonlinear response in a narrow band (emitted photons between$\sim$1.47 and 1.63eV), but no resonant behavior nor doping dependence. Ref.\citenum{AlexCondMat2017} reported a factor$\sim$2 enhancement of the third order nonlinear signal in a FWM experiment at the onset of inter-band transitions ($\hbar\omega_{0}$=2$|E_{\rm F}|$, where $E_{\rm F}$ is the Fermi Energy) for SLG on SiN waveguides in a narrow band (emitted photon energies between$\sim$0.79 and 0.8eV). Thus, to date, evidence of tunable third order nonlinear effects in SLG is limited to narrow bands and weak enhancements.

Here we show that THGE in SLG can be tuned by almost two orders of magnitude over a broad energy range (emitted photon energies between$\sim$1.2-2.2eV) and over 20 times by electrical gate tuning. These results, in agreement with calculations based on the intrinsic third-harmonic conductivity of massless Dirac fermions, confirm that SLG is a unique nonlinear material since it allows electrical tuning of $\chi^{(3)}$ over an ultra-broad range, only limited by the linearity of the Dirac cone  ($\pm 2$eV\cite{NetoRevModPhys2009}).

In order to test both the photon energy dependence and the electrical tunability of THGE we use two sets of samples: SLG placed on a transparent substrate (sapphire, Sa), Fig.\ref{fig:samples}a, and back-gated SLG on a reflective substrate (Si/SiO$_2$), Fig.\ref{fig:samples}b. To study the THGE photon energy dependence, we measure it over a broad range (incident photon energy$\sim$0.4-0.7eV, with a THG signal at$\sim$1.2-2.1eV only limited by the absorption of the Si based charge-coupled device, CCD, used in our set-up). Transmission measurements allow us to derive the absolute THGE, by taking into account the system losses and by minimizing the chromatic aberrations of the optical components (e.g. in reflection one needs to use beam splitters and these do not have a flat response over the$\sim$0.4-2.2eV range). Thus, we use chemical vapor deposition (CVD) to obtain a large area SLG ($\sim$cm size) and simplify the alignment, given the low optical contrast of SLG on sapphire\cite{CasiNL2007}. When measuring the THGE electrical tunability we need to follow the THG intensity normalized to its minimum, as function of gate voltage ($V_G$). For each $\omega_0$ we measure 62 spectra, one for each $V_G$. For each spectrum we calculate the total number of counts on the CCD, which is proportional to the total number of THG photons, and divide all the spectra by that with the minimum counts. The key here is a precise control of $E_F$, while any system uncertainties on the absolute THGE are removed by the normalization. Thus, we use an exfoliated SLG back-gated field-effect transistor (FET) on Si+285nm SiO$_{2}$.

The two sets of samples are prepared and characterized as described in Methods (Sect.\ref{subs:samples}). $E_F$ for the CVD SLG on Sa is$\sim$250meV, and $<$100meV in the exfoliated SLG on Si+285nm SiO$_2$. The defect density is n$_{\rm D}\sim$6$\times$10$^{10}$cm$^{-2}$ for SLG on Sa and n$_{\rm D}\sim$2.4$\times$10$^{10}$cm$^{-2}$ for SLG on Si/SiO$_2$. The different $E_F$ is considered in our theory and addressed experimentally, since $E_F$ is one of the parameters of our study. The small difference in defects suggests that the two samples are comparable. In fact, as we discuss in the following, THGE has a negligible dependence on disorder, impurities and imperfections over a range of values that covers the vast majority of SLG reported in literature.
\begin{figure}
\centerline{\includegraphics[width=90mm]{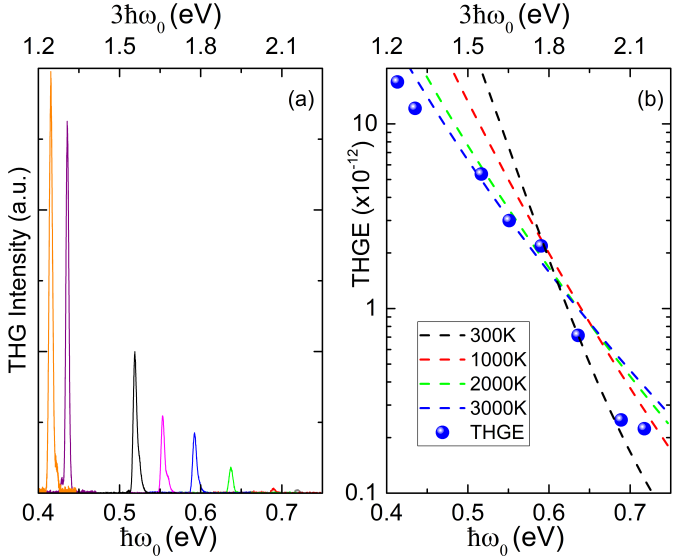}}
\caption{\textbf{Energy dependence of SLG THGE} (a) THG spectra for $\hbar\omega_{0}\sim0.4$ to$\sim0.7$eV and an average power$\sim$1mW. (b) THGE for SLG on Sa as a function of $\hbar\omega_{0}$ (x bottom axis) and $3\hbar\omega_{0}$ (x top axis). Curves are calculated for $\tau \gg$ 1 ps and increasing $T_e$ for $E_F$=250meV and $I_{\omega_{0}}\sim 2.4\times 10^{12}$Wm$^{-2}$.}
\label{fig:THG_energy}
\end{figure}
\begin{figure}
\centerline{\includegraphics[width=90mm]{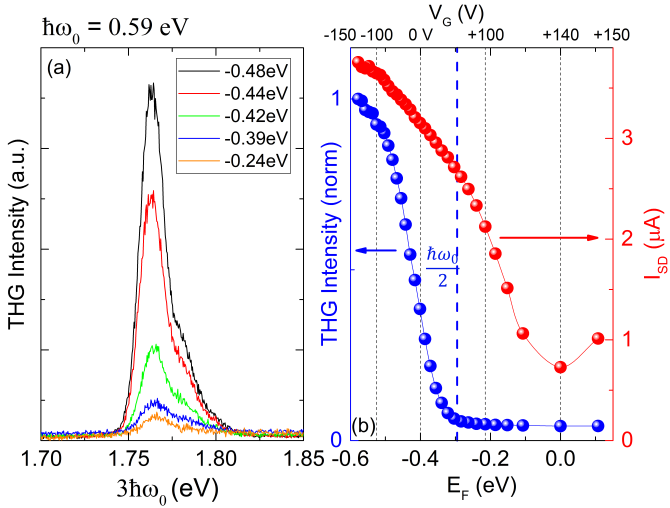}}
\caption{\textbf{Gate tunability of THG} (a) THG spectrum as a function of $E_F$ for exfoliated SLG on Si/SiO$_2$. (b) THG intensity (left y axis, blue dots) and $I_{SD}$ (right y axis, red dots) as a function of $E_F$ (bottom x axis) and corresponding $V_G$ (top x axis) for SLG on Si/SiO$_2$.}
\label{fig:THG_gate}
\end{figure}
\begin{figure*}[t]
\centerline{\includegraphics[width=180mm]{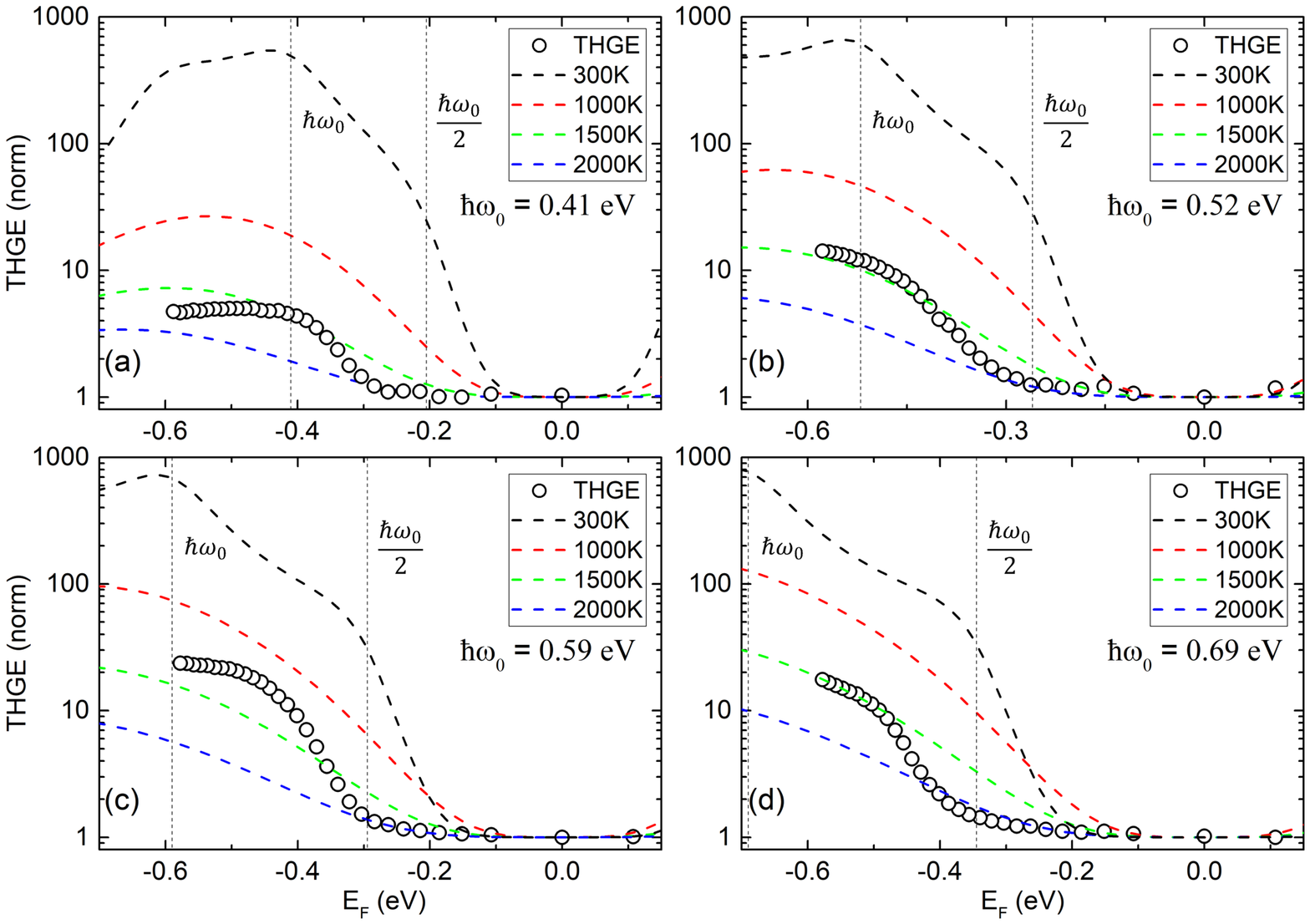}}
\caption{\textbf{Broadband THGE electrical modulation.} Experiments (circles) and theory (dotted lines) for THGE as a function of $E_F$ and for different $T_e$ for SLG on Si/SiO$_2$ and incident photon energies of (a)0.41, (b)0.52, (c)0.59, (d)0.69eV. The vertical dotted lines in each panel are at $|E_F| = \hbar\omega_0/2$ and $|E_F|=\hbar\omega_0$.}
\label{fig:THG_TheoryExp}
\end{figure*}

The THGE measurements are performed in air at room temperature for both transmission and reflection, as detailed in Methods (Sect.\ref{subs:setup}). Fig.\ref{fig:THG_energy}a plots representative TH spectra for different incident $\hbar\omega_0$ for SLG on Sa. We assign the measured signal to THG because the energy of the detected photons is equal to three times that of the incident one: $\hbar\omega_{THG}=3\hbar\omega_0$ and its intensity scales with the cube of the incident power ($I_{3\omega_0} \propto I^3_{\omega_0}$)\cite{Boyd2003}. Fig.\ref{fig:THG_energy}b shows that when $\hbar\omega_0$ decreases from$\sim0.7$ to$\sim 0.4$eV THGE is enhanced by a factor$\sim$75, almost one order of magnitude larger than the$\sim9.5$ expected from the $1/\omega_0^4$ dependence of THGE in SLG at $E_F$=0\cite{RostPRB2016}. This can be explained by taking into account the dependence of the SLG third order optical conductivity $\sigma^{(3)}_{\ell\alpha_1\alpha_2\alpha_3}$, where $\ell,\alpha_{i=1,2,3}$ are the Cartesian indexes, on $\omega_0$, $E_F$ and electronic temperature ($T_e$). Note that $\chi^{(3)} \equiv i \sigma^{(3)}/(3\epsilon_0\omega_0 d_{eff})$\cite{ChengNJPhys2014}, by considering SLG with a thickness $d_{eff}$. Our modeling is based on $\sigma^{(3)}_{\ell\alpha_1\alpha_2\alpha_3}$, therefore we do not need to use $d_{eff}$.

The dependence on $E_F$ was calculated in Refs.\cite{MikhPRB2016,RostPRB2016} and gives resonances at $\hbar\omega_0=2|E_F|,|E_F|,2|E_F|/3$ for $T_e$=0K. A finite $T_e$ modifies the height and broadening of these resonances, as derived in Methods (Sect.\ref{subs:theory}). A comparison between theoretical curves, for $E_F$=250meV and different $T_e$, and experiments is plotted in Fig.\ref{fig:THG_energy}b. This indicates that $E_F$ plays a key role, in particular when $\hbar\omega_0\leq$2$|E_F|$. The effects of disorder, impurities and imperfections can be phenomenologically introduced by a relaxation rate, $\Gamma=\hbar/\tau$, through the density matrix approach\cite{ChengPRB2015,MikhPRB2016}. Our analysis (see Methods Sect.M3.3) shows that the effect of a finite $\tau $ in the $\sim$0.1fs-1ps range on THGE is negligible. Since $\Gamma\sim e\hbar v^2_{\rm F}/(\mu_{e}E_{\rm F})$\cite{ref:proof} this range of $\tau$, for SLG with $E_F$ between 100 and 600meV, would correspond to mobilities$\sim$1-10$^5$ cm$^2$V$^{-1}$s$^{-1}$, covering the vast majority of experimental SLG in literature.

Refs.\citenum{MikhPRB2016,RostPRB2016} predicted that gate tunability of THGE should be possible. Fig.\ref{fig:THG_gate}a plots the THG spectra for different V$_G$ and $\hbar\omega_0$=0.59eV. Fig.\ref{fig:THG_gate}b shows the THG intensity over -600meV$\leq E_{\rm F} \leq$+150meV corresponding to-150V$\leq V_G \leq$+150V. $E_F$ is derived from each $V_G$ as discussed in Methods Sect.M1.

Fig.\ref{fig:THG_gate}b shows that, as a function of $V_G$, there is a THG intensity enhancement by over a factor of 20 when $\hbar\omega_0<2|E_F|$. Fig.\ref{fig:THG_gate}b also indicates that THGE in SLG follows an opposite trend compared to FWM\cite{AlexCondMat2017}: it is higher for intra-band ($\hbar\omega_0<2|E_F|$) than inter-band ($\hbar\omega_0>2|E_F|$) transitions. This is reproduced by the calculations in Methods (Sect.\ref{subs:theory}). THGE for SLG, considered as a nonlinear interface layer between air and substrate, under normal incidence can be written as (see Methods Sect.M3.1):
\begin{equation}\label{eq:THG_efficiency}
\eta^{THG}(\omega_0,E_F,T_e)=\frac{I_{3\omega_0}}{I_{\omega_0}}=f(\omega_0)\frac{I^2_{\omega_0}}{4\epsilon^4_0 c^4}\left|\sigma^{(3)}_{\ell\ell\ell\ell}(\omega_0,E_F,T_e)\right |^2
\end{equation}
where $\epsilon_0\sim 8.85\times 10^{-12}{\rm C (V m)^{-1}}$ and $c=3\times10^8$m/s are the vacuum permittivity and the speed of light; $f(\omega_0)=n^{-3}_1(\omega_0)n_2(3\omega_0)[n_1(3\omega_0)+n_2(3\omega_0)]^{-2}$ in which $n_{i=1,2}(\omega)$ is the refractive index of air (i=1) and substrate (i=2). For SLG on any substrate $n_1({\omega})\sim1$ and $n_2({\omega})=\sqrt{\epsilon_{2}(\omega)}$, with $\epsilon_{2}(\omega)$ the substrate dielectric function. For Sa, we use $\epsilon_2\sim10$\cite{SaSpecs}. According to the $C_{6v}$ point group symmetry of SLG on a substrate\cite{RostNC2017}, the relative angle between laser polarization and SLG lattice is not important for the third-order response (see Methods Sect.M3.2). We can thus assume the incident light polarization $\hat{\ell}$ along the zigzag direction without loss of generality. $\sigma^{(3)}_{\ell\ell\ell\ell}$ can then be calculated employing a diagrammatic technique\cite{RostPRB2016,RostPRB2017,RostNC2017}, where we evaluate a four-leg Feynman diagram for the TH response function (see Methods Sect.\ref{subs:theory}). The light-matter interaction is considered in a scalar potential gauge in order to capture all intra-, inter-band and mixed transitions\cite{ChengNJPhys2014,ChengPRB2015,MikhPRB2016,RostPRB2016}.

Fig.\ref{fig:THG_TheoryExp} compares experiments and theory for THGE for four incident photon energies between 0.41 and 0.69eV and different T$_e$. Both theory and experiments display a plateau-like feature for THGE at low $E_F$ ($2|E_F|<\hbar\omega_0$), which corresponds to inter-band transitions for the incident photon. By further increasing $|E_F|$, we reach the energy region for intra-band transitions ($\hbar\omega_0<2|E_F|$), where we observe a THGE rise up to a maximum for $|E_F|\sim1.25\hbar\omega_0$ (Fig.\ref{fig:THG_TheoryExp}a). This is due to the merging of the two $T_e$=0K resonances at $|E_F|/\hbar\omega_0=$1 and 1.5 as a result of high $T_e$ (see Methods Sect.M3.4). Fig.\ref{fig:THG_TheoryExp} indicates that the best agreement between theory and experiments is reached when $\sim$1500K$\leq T_e\leq$2000K.

\begin{figure}
\centerline{\includegraphics[width=80mm]{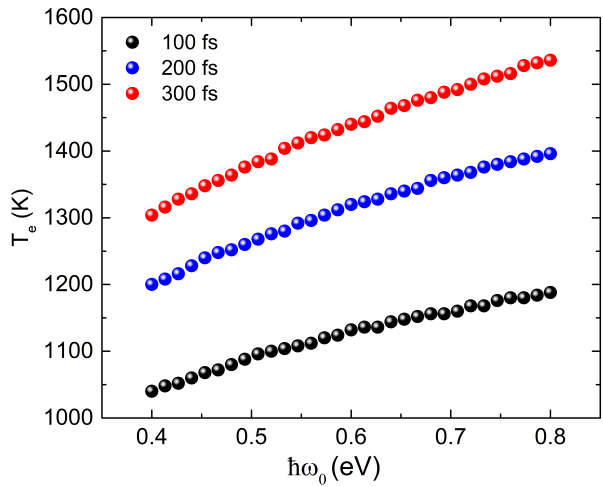}}
\caption{\label{fig:steady_state_temperature}
\textbf{Steady-state T$_e$ in photoexcited SLG}. $T_e$ as a function of $\hbar\omega_0$ for $\tau$=100 (black), 200 (blue), and 300fs (red).}
\end{figure}

$T_e$ can be also independently estimated as follows. When a pulse of duration $\Delta t$ and fluence ${\cal F}$, with average absorbed power per unit area P/A, photoexcites SLG, the variation dU of the energy density in a time interval dt is dU=(P/A)dt. The corresponding $T_e$ increase is dT$_e$=dU/c$_v$, where $c_v$ is the electronic heat capacity of the photoexcited SLG, as derived in Methods (Sect.\ref{subs:Te}). When the pulse is off, $T_e$ relaxes towards the lattice temperature on a time-scale $\tau$. This reduces $T_e$ by $dT_e=-(T_e/\tau)dt$ in a time interval dt. Thus:
\begin{equation}
\frac{dT_{\rm e}}{dt}=\frac{1}{c_{\rm v}}\frac{P}{A}-\frac{T_{\rm e}}{\tau}~.
\end{equation}
If the pulse duration is: (i) much longer than$\sim$20fs, which is the time-scale for the electron distribution to relax to the Fermi-Dirac profile in both bands\cite{BridNC2013}; (ii) comparable to the time-scale$\sim150-200fs$ needed to heat the optical phonon modes\cite{BridNC2013,LazzPRL2005,PiscPRL2004}, it is safe to assume that $T_e$ reaches a steady-state during the pulse, given by:
\begin{equation}\label{eq:steady_temp}
T_e= \frac{\tau}{c_v(\mu_c,\mu_v, T_e)}\frac{P}{A}~.
\end{equation}
The $T_e$ dependence of $c_v$ in Eq.\ref{eq:steady_temp} is discussed in Methods (Sect.\ref{subs:Te}). In our experiments we have: ${\cal F}$=70$\mu$J/cm$^2$, $\Delta$t=300fs,  P/A=2.3\%$\times {\cal F}/\Delta t$. The resulting $T_e$, as a function of $\hbar\omega_0$, for several $\tau$, are in Fig.\ref{fig:steady_state_temperature}. $T_e$ increases for more energetic photons and for longer $\tau$. Overall, $T_e$ ranges between$\simeq$1000 and 1500K, in excellent agreement with the estimate from Fig.\ref{fig:THG_TheoryExp}.
\begin{figure}
\centerline{\includegraphics[width=80mm]{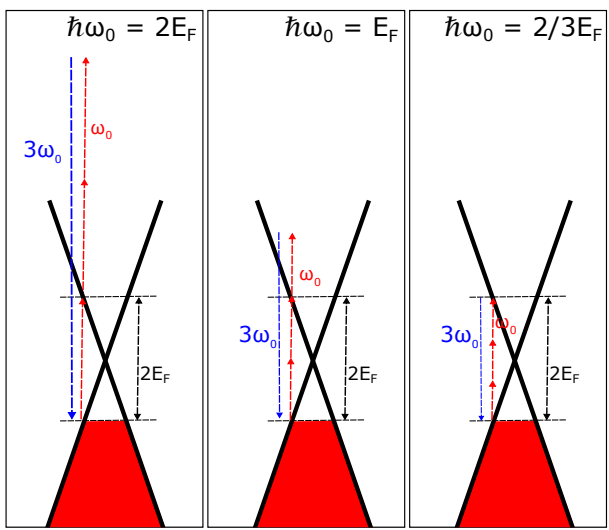}}
\caption{\textbf{Multi-photon resonances in SLG.} Resonances corresponding to the three logarithmic peaks in the imaginary part of the SLG nonlinear conductivity that occur at $T_e$=0K for $\hbar \omega_0=2|E_F|/3$, $|E_F|$, $3|E_F|/2$. The red arrows represent the incident $\omega_{0}$ photons and the blue arrows represent the TH photons at $3\omega_0$.}
\label{fig:THG_sketch_resonances}
\end{figure}

The observed gate-dependent enhancement of the THGE can be qualitatively understood as follows. The linear optical response of SLG at T$_e$=0K has a "resonance" for $\hbar\omega_0=2|E_F|$, the onset of intra- and inter-band transitions\cite{MakPRL2008}. Around this energy, a jump occurs in the real part of $\sigma^{(3)}_{\ell\alpha_1\alpha_2\alpha_3}$ due to the relaxation of the Pauli blocking constraint for vertical transitions between massless Dirac bands. From the Kramers-Kronig relations\cite{KronJOSA1926}, this jump corresponds to a logarithmic peak in the imaginary part of $\sigma^{(3)}_{\ell\alpha_1\alpha_2\alpha_3}$. In a similar way, for the SLG third-order nonlinear optical response, logarithmic peaks in the imaginary part of $\sigma^{(3)}_{\ell\alpha_1\alpha_2\alpha_3}$ occur at T$_e$=0K for multi-photon transitions such that $m \hbar \omega = 2|E_F|$ with $m$=1,2,3, which correspond to incident photon energies $\hbar \omega_{0}=2|E_F|$, $|E_F|$, 2/3$|E_F|$\cite{MikhPRB2016,RostPRB2016}, as sketched in Fig.\ref{fig:THG_sketch_resonances}. At high T$_e$, due to the broadening of the Fermi-Dirac distribution, these peaks are smeared and merge (see Methods Sect.M3.4). Our work provides experimental evidence of this resonant structure (Fig.\ref{fig:THG_TheoryExp}), in agreement with theory.

In summary, we demonstrated that the THG efficiency in SLG can be modulated by over one order of magnitude by controlling its $E_F$ and by almost two orders of magnitude by tuning the incident photon energy in the range$\sim$0.4-0.7eV. The observation of a steep increase of THGE at $|E_F|$=$\hbar\omega_0/2$ for all the investigated photon energies suggests that the effect can be observed over the entire linear bandwidth of the SLG massless Dirac fermions. These results pave the way to novel SLG-based nonlinear photonic devices, in which the gate tuneability of THG may be exploited to implement on-chip schemes for optical communications and signal processing, such as ultra-broadband frequency converters.
\section*{\label{Meth}Methods}
\subsection{Sample preparation and characterization}
\label{subs:samples}
SLG on Sa is prepared as follows. SLG is grown by CVD on Cu as for Ref.\cite{LiS2009}. A Cu foil (99.8\% pure) substrate is placed in a furnace. Annealing is performed at 1000 $^{\circ}$C in a 20sccm (standard cubic centimeters per minute) hydrogen atmosphere at $\sim$196mTorr for 30min. Growth is then initiated by introducing 5sccm methane for 30mins. The grown film is characterized by Raman spectroscopy\cite{FerrPRL2006,FerrNN2013} with a Horiba Labram HR800 spectrometer equipped with a 100x objective at 514nm, with a power on the sample$\sim$500$\mu$W to avoid any possible heating effects. The D to G intensity ratio is I(D)/I(G)$\ll$0.1, corresponding to a defect density n$_{\rm D}\ll2.4\times10^{10}$cm$^{-2}$\cite{CancNL2011,BrunaACSNano2014}. The 2D peak position (Pos) and full width at half maximum (FWHM) are Pos(2D)$\sim$2703cm$^{-1}$ and FWHM(2D)$\sim$36cm$^{-1}$, respectively, while Pos(G)$\sim$1585cm$^{-1}$ and FWHM(G)$\sim$18cm$^{-1}$. The 2D to G intensity and area ratios are I(2D)/I(G)$\sim$3.3 and A(2D)/A(G)$\sim$6.5, respectively. The CVD SLG is then transferred on Sa by polymer-assisted Cu wet etching\cite{BonaMT2012}, using polymethyl methacrylate (PMMA). After transfer Pos(2D)$\sim$2684cm$^{-1}$, FWHM(2D)$\sim$24cm$^{-1}$, Pos(G)$\sim$1584cm$^{-1}$, FWHM(G)$\sim$13cm$^{-1}$, I(2D)/I(G)$\sim$5.3, A(2D)/A(G)$\sim$10. From Refs.\citenum{BaskPRB2009,DasNN2008} we estimate $E_F\sim$250meV. After transfer, I(D)/I(G)$\sim$0.14, which corresponds to $n_{\rm D}\ll6.0\times10^{10}$cm$^{-2}$\cite{CancNL2011,BrunaACSNano2014} with a small increase of defect density.

The back-gated SLG sample is prepared by micromechanical exfoliation of graphite on Si+285nm SiO$_2$\cite{NovoPNAS2005}. Suitable single-layer flakes are identified by optical microscopy\cite{CasiNL2007} and Raman spectroscopy\cite{FerrPRL2006,FerrNN2013}. The device is then prepared as follows. We deposit a resist (A4-495) on the exfoliated SLG on Si/SiO$_2$ and we pattern it with electron beam lithography. Then, we develop the resist in a solution of isopropanol (IPA) diluted with distilled water, evaporate and lift-off 5/70nm of Cr/Au. Cr is used to improve adhesion of the Au, while Au is the metal for source-drain contacts. Raman spectroscopy is then performed after processing. We get Pos(2D)$\sim$2678cm$^{-1}$, FWHM(2D)$\sim$25cm$^{-1}$, Pos(G)$\sim$1581cm$^{-1}$, FWHM(G)$\sim$12cm$^{-1}$, I(2D)/I(G)$\sim$4.9, A(2D)/A(G)$\sim$10.3, indicating E$_F<$100meV\cite{BaskPRB2009,DasNN2008}. I(D)/I(G)$\ll$0.1, corresponding to n$_D\ll2.4\times10^{10}$cm$^{-2}$\cite{CancNL2011,BrunaACSNano2014}. When V$_G$ is applied, $E_{\rm F}$ is derived from V$_G$ as follows\cite{DasNN2008}:
\begin{equation} \label{eq:Ef}
E_{\rm F} = \hbar v_{\rm F} \sqrt{\pi n}
\end{equation}
where $\hbar$ is the reduced Plank constant, and $n$ is the SLG carrier concentration. This can be written as\cite{DasNN2008}:
\begin{equation} \label{eq:nFET}
n = \frac{C_{\rm BG}}{e} (V_{\rm G}-V_0)
\end{equation}
where $C_{BG}=\epsilon\epsilon_0/d_{BG}=1.2 \times 10^{-8}$Fcm$^{-2}$ is the back-gate capacitance (d$_{BG}$=285nm is the back-gate thickness and $\epsilon\sim$4 the SiO$_2$ dielectric constant\cite{DasNN2008}), $e>0$ is the fundamental charge and $V_0$ is the voltage at which the resistance of the SLG back-gated device reaches its maximum (minimum of the current between source and drain). We note that the SLG quantum capacitance (C$_{QC}$) is negligible in this context. In fact the SiO$_2$ layer and SLG can be considered as two capacitors in series and the SLG C$_{QC}$ is $\sim$10$^{-6}$Fcm$^{-2}$\cite{XiaNN2009}. Thus the total capacitance C$_{tot}=$(1/C$_{BG}$+1/C$_{QC}$)$^{-1}\sim$C$_{BG}$.
\subsection{THGE measurements and calibration}
\label{subs:setup}
\begin{figure}
\centerline{\includegraphics[width=90mm]{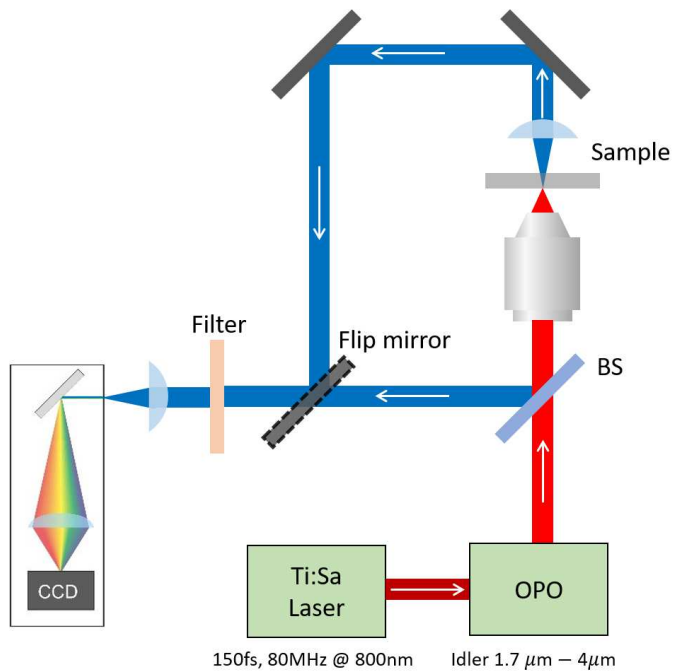}}
\caption{Setup used for THG experiments in both transmission and reflection. BS, Beam Splitter.}
\label{fig:setup}
\end{figure}
THGE measurements are performed in air at room temperature for both transmission and reflection, as shown in Fig.\ref{fig:setup}. For excitation we use the idler beam of an OPO (Coherent) tuneable between$\sim$0.31eV (4$\mu$m) and$\sim$0.73eV (1.7$\mu$m). This is seeded by a mode-locked Ti:Sa laser (Coherent) with 150fs pulse duration, 80MHz repetition rate and 4W average power at 800nm. The OPO idler is focused by a 40X reflective objective (Ag coating, numerical aperture NA=0.5) to avoid chromatic aberrations. The THG signal is collected by the same objective (in reflection mode) or collimated by an 8mm lens (in transmission mode) and delivered to a spectrometer (Horiba iHR550) equipped with a nitrogen cooled Si CCD, Fig.\ref{fig:setup}. The idler spot-size is measured with the razor-blade technique\cite{DomiEL1999} to be$\sim$4.7$\mu$m. This corresponds to an excitation fluence$\sim$70$\mu$J/cm$^{2}$ for the average power (1mW) used in our experiments. The idler pulse duration is checked by autocorrelation measurements based on two-photon absorption on a single channel Si photodetector and is$\sim$300fs. Under these excitation conditions, the THG signal is stable over at least 1 hour. For the electrical dependent THGE measurements we use a Keithley 2612B dual channel source meter to apply $V_G$ between -150 and +150V, a source-drain voltage (10mV), as well as to read the source-drain current ($I_{SD}$). For the photon energy dependence measurements we use 60s acquisition time and 10 accumulations (giving a total of 10 minutes for each spectrum). For the gate dependence measurements we proceeded as follows. We tune $V_{\rm G}$ (62 points between -150 and +150V) and for each $V_G$ we measure the THG signal by using 10s acquisition time and 1 accumulation. We use a shorter accumulation time compared to the photon energy measurements to reduce the total time required for each $V_{\rm G}$ scan. A lower accumulation time implies that less photons are collected by the CCD. We consider the amplitude of THG in counts/s, obtained by dividing the number of counts detected on the CCD by the accumulation time. Thus, in the case of $V_G$ dependent measurements, SLG is kept at a given $V_G$ for 10s before moving to the next point (next value of $V_G$). This corresponds to$\sim$10minutes for each measurement (\emph{i.e.} a full $V_G$ scan between -150 and + 150V). In this way, for each $V_G$ and, consequently, for each $E_F$, we record one THG signal spectrum.

To estimate the $\omega_0$ dependent THGE, it is necessary to first characterize the photon energy dependent losses of the optical setup. The pump-power is measured on the sample (by removing it and measuring the power after the objective). The major losses along the optical path are the absorption of Sa, the grating efficiency, and the CCD quantum efficiency. We also need to consider the CCD gain. The Sa transmittance is$\sim$85\% in the energy range of our THG experiments\cite{SaSpecs}. To evaluate the losses of the grating and the absorption of the CCD, we align the Ti:Sa laser, tuneable between$\sim$1.2 and 1.9eV ($\sim$650-1050nm), with the microscope and detect it on the CCD. We then measure the signal on the spectrometer, given a constant number of photons for all wavelengths, and compare this with the spectrometer specifications\cite{SymphonySpecs}. We get an excellent agreement between the two methods (\emph{i.e.} evaluation of the losses from detection of the fundamental beam and spectrometer specifications). Thus we use the spectrometer specifications to estimate the losses due to grating and CCD efficiencies. We also account for the CCD gain, \emph{i.e.} the number of electrons necessary to have 1 count. The instrument specs\cite{SymphonySpecs} give a gain$\sim$7.
\subsection{TGHE modeling}
\label{subs:theory}
\begin{figure}
\centerline{\includegraphics[width=50mm]{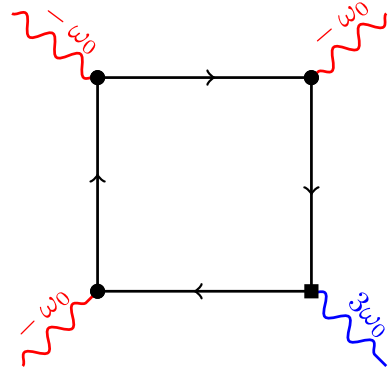}}
\caption{Feynamn diagram for $\Pi^{(3)}_{\ell}$ in the scalar potential gauge. Solid/wavy lines indicate non-interacting Fermionic propagators/external photons. Solid circles and square indicate density and current vertexes}
\label{fig:feyn}
\end{figure}

$\sigma^{(3)}_{\ell\ell\ell\ell}$ is calculated through a diagrammatic technique, with the light-matter interaction taken in the scalar potential gauge in order to capture all intra-, inter-band and mixed transitions\cite{ChengPRB2015,MikhPRB2016,RostPRB2016}. We evaluate the diagram in Fig.\ref{fig:feyn} and denote by $\Pi^{(3)}_{\ell}$ the response function. $\hat{n}$ and $\hat{j}_\ell$ are the density and paramagnetic current operators. Then, $\sigma^{(3)}_{\ell\ell\ell\ell}=(i e)^3 \lim_{{\vec q}\to {  0}} \partial^3\Pi^{(3)}_{\ell}/\partial q^3_{\ell}$, where $e>0$ is the fundamental charge\cite{RostPRB2016}. We use the Dirac Hamiltonian of low-energy carriers in SLG as ${\cal H}_{k} = \hbar v_{\rm F} {\vec \sigma}\cdot{\vec k}$ where ${ \vec \sigma} = (\tau\sigma_x,\sigma_y)$ stands for the Pauli matrices in the sublattice basis. Note that $\tau=\pm$ stands for two valleys in the SLG Brillouin zone. We get $\sigma^{(3)}_{xxxx}(\omega,E_{\rm F},0)=i\sigma^{(3)}_{0} \bar{\sigma}^{(3)}_{xxxx}(\omega,E_{\rm F},0)$ at $T_{\rm e}=0$\cite{ChengPRB2015,MikhPRB2016,RostPRB2016}:
\begin{align}\label{eq:clean}
\bar{\sigma}^{(3)}_{xxxx} (\omega,E_{\rm F},0) &=\frac{ 17  G(2|E_{\rm F}|,\hbar\omega_{+})  - 64 G(2|E_{\rm F}|,2\hbar\omega_{+})|}{24 (\hbar\omega_{+})^4}
\nonumber\\  &+ \frac{45 G(2|E_{\rm F}|,3\hbar\omega_{+})}{24 (\hbar\omega_{+})^4}
\end{align}
where $G(x,y)= \ln|(x+y)/(x-y)|$, $\sigma^{(3)}_0 ={N_{\rm f} e^4 \hbar v^2_{\rm F}}/({32\pi})$ with $N_{\rm f}=4$ and $\hbar\omega_{+} \equiv  \hbar\omega+i0^+$. At finite $T_{\rm e}$, $\sigma^{(3)}_{\ell\ell\ell\ell}$ is evaluated as\cite{Vignale_book}:
\begin{equation}
\sigma^{(3)}_{xxxx}(\omega,E_{\rm F},T_{\rm e}) =\frac{1}{4k_{B}T_{\rm e}} \int^{\infty}_{-\infty}dE~\frac{ \sigma^{(3)}_{xxxx}(\omega,E,0) }{  \cosh^2  \left (\frac{E-\mu}{2k_{B}T_{\rm e}} \right ) }.
\end{equation}
\begin{figure}[htp]
\centerline{\includegraphics[width=80mm]{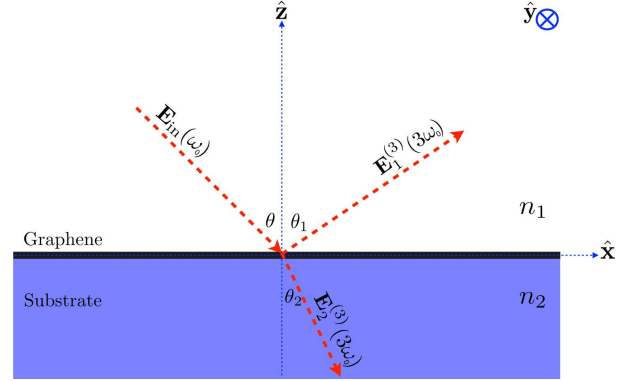}}
\caption{Schematic of SLG on substrate. The TH radiated waves in the top and bottom medium obey the TH Snell's law: $n_i(3\omega_0)\sin\theta_i=n_1(\omega_0)\sin\theta$. The red dashed arrows indicate the propagation direction of in-coming and out-going waves.}
\label{fig:snell}
\end{figure}
\subsubsection{THGE of SLG as an interface layer}
\label{subss:THGElayer}
In order to evaluate the THGE for SLG on a substrate we consider SLG as an interface layer between air and substrate, see Fig.\ref{fig:snell}, and implement electromagnetic boundary conditions for the non-harmonic radiations. We follow Ref.\citenum{ShenARPC1989}, and provide explicit details for THG in SLG. The Maxwell equations in the nonlinear medium in the $m(\ge 2)$-th order of perturbation are given by\cite{Boyd2003,Jackson_book}:
\begin{align}
 &{\vec \nabla}\cdot {\vec B}^{(m)} = 0~,\\
 & {\vec \nabla}\cdot {\vec D}^{(m)} = \frac{\rho^{(m)}_{\rm f}}{\epsilon_0} -\frac{1}{\epsilon_0}{\vec \nabla}\cdot {\vec P}^{(m)}~, \\
 &{\vec \nabla}\times {\vec E}^{(m)} = i \omega_\Sigma   {\vec B}^{(m)} ~, \\
 &{\vec \nabla}\times {\vec B}^{(m)} = \mu_0 {\vec J}^{(m)}_{\rm f} -i\frac{\omega_\Sigma}{c^2} {\vec D}^{(m)} -i\omega_\Sigma \mu_0 {\vec P}^{(m)}~.
\end{align}
where ${\vec D}^{(m)}=\epsilon (\omega_\Sigma)  {\vec E}^{(m)}$ is the {\it conventional} displacement vector. $\rho^{(m)}_{\rm f}$ and ${\vec J}^{(m)}_{\rm f}$ are the $m$-th order Fourier components of free charge and free current. Note that $\omega_\Sigma=\sum^m_{i}\omega_i$ in which $\omega_i$ correspond to the incoming photons frequency into the nonlinear medium. In the case of THG, we have $m=3$, $\omega_{1,2,3}=\omega_0$ and $\omega_\Sigma=\omega_{\rm THG}=3\omega_0$. $\epsilon(\omega)$ is the isotropic and homogenous linear relative dielectric function. Only electric-dipole contributions are included.

We consider SLG in the $x$-$y$ plane embedded between air and a substrate. SLG is modeled by a dielectric function $\epsilon_{\rm s}(\omega)$, nonlinear polarization, free surface charge and free surface current:
\begin{align}
{\vec P}^{(m)} &= \delta(z) {\vec {\mathcal P}}^{(m)}~, \\
\rho^{(m)}_{\rm f} &=\delta(z)\sigma^{(m)}_{\rm f}~,\\
 {\vec J}^{(m)}_{\rm f} &= \delta(z) {\vec K}^{(m)}_{\rm f}~.
\end{align}
Having the Dirac delta, $\delta(z)$, in the above relations implies that SLG only shows up in the electromagnetic boundary conditions. Note that ${\vec  {\mathcal P}}^{(m)}$ and $ {\vec K}^{(m)}_{\rm f}$ are in-plane vectors with zero component along the interface normal, $\hat  z$. The interface layer is the only source of nonlinearity. We assume $\sigma^{(m)}_{\rm f}=0$ and ${\vec K}^{(m)}_{\rm f}={0}$, consistent with our experiments, where there are no free surface charges and currents that oscillate at frequency $m\omega$ with $m=2,3,\dots$.

The boundary conditions for the nonlinear fields at z=0 are obtained as:
\begin{align}\label{eq:BC}
&~{\vec B}^{(m)}_{1} - {\vec B}^{(m)}_{2}  = \mu_0 ({\vec K}^{(m)}_{\rm f}-i\omega_\Sigma {\vec  {\mathcal P}}^{(m)}) \times \hat  {z} ~, \nonumber \\
&\left \{ \epsilon_{1}(\omega_\Sigma){\vec E}^{(m)}_{1} - \epsilon_{2}(\omega_\Sigma){\vec E}^{(m)}_{2}
\right \} \cdot \hat { z} = \frac{\sigma^{(m)}_{\rm f}- {\vec \nabla}_{\rm 2D}\cdot {\vec {\mathcal P}}^{(m)}}{\epsilon_0} ~,\nonumber\\
 &~({\vec E}^{(m)}_{1} - {\vec E}^{(m)}_{2})\times \hat{ z} = { 0} ~.
\end{align}
Where the sub-indexes 1,2 stand for the top(bottom) medium and ${\vec \nabla}_{\rm 2D}= \hat{ x}\partial /\partial x+ \hat{ y}\partial /\partial y$. The dielectric function of the interface layer, $\epsilon_{\rm s}(\omega)$, does not emerge in the above boundary conditions.

The wave equation in the top and bottom media, with vanishing nonlinear polarization, follows:
\begin{equation}
{\vec \nabla} \times {\vec \nabla} \times {\vec E}^{(m)}  - \ \frac{\omega^2_\Sigma}{c^2}   \epsilon(\omega_\Sigma)   {\vec E}^{(m)} =  0~.
\end{equation}
which has a plane wave solution\cite{Jackson_book}:
\begin{align}
 {\vec E}^{(m)} = \hat{ \ell} {\cal E}^{(m)} e^{i({\vec q}_{\Sigma}  \cdot {\vec r} -\omega_\Sigma t)} +c.c.
\end{align}
$\hat{ \ell} \cdot {\vec q}_{\Sigma} = 0$ and the dispersion relation in the top and bottom media is given by:
\begin{align}
q_{\Sigma} =\left |{\vec q}_{\Sigma} \right |=  \frac{\omega_\Sigma}{c} n(\omega_{\Sigma})~.
\end{align}
where $n(\omega_\Sigma) =\sqrt{\epsilon(\omega_\Sigma)} $ is the refractive index of the lossless media.

We consider a linearly polarized incident laser with arbitrary incident angle exposed to the interface layer:
\begin{align}
{\vec E}_{\rm in} =\left\{ \hat{  x} {\cal E}_{x} + \hat{  y} {\cal E}_{y}+ \hat{  z} {\cal E}_{z} \right\} e^{i({\vec q}\cdot {\vec r} - \omega_0 t)} +c.c.
\end{align}
where
\begin{align}
{\vec q}= \frac{\omega_0}{c} n_{1}(\omega_0) [-\cos\theta \hat{  z}+\sin\theta \hat{  x}]~.
\end{align}
The leading nonlinearity of SLG is encoded in a third-order conductivity tensor, $\overset\leftrightarrow {\sigma}^{(3)}$. Using the SLG symmetry, the third-order nonlinear polarization follows:
\begin{align}
{\vec {\mathcal P}}^{(3)} =\vec {\widetilde{ \mathcal {P}}}^{(3)} \exp \left\{ i\frac{3\omega_0}{c} \left [n_{1}(\omega_0) x \sin\theta  -c t \right ]\right\} +c.c.
\end{align}
where
\begin{align}\label{eq:p3}
&\widetilde{\cal P}^{(3)}_x =\frac{i}{3\omega_0}\sigma^{(3)}_{xxxx} \left \{ {\cal E}^3_{x} +{\cal E}_{x} {\cal E}^2_{y}\right \}~, \nonumber \\
&\widetilde{\cal P}^{(3)}_y =\frac{i}{3\omega_0}\sigma^{(3)}_{xxxx} \left \{    {\cal E}^3_{y} + {\cal E}_{y} {\cal E}^2_{x}\right \} ~, \nonumber \\
&\widetilde{\cal P}^{(3)}_z=0~.
\end{align}
The wave-vectors of TH radiated waves in the top and bottom media are then:
\begin{align}\label{eq:q3tb}
&{\vec q}_{3\omega_0,1}=\frac{3\omega_0}{c} n_{1}(3\omega_0) [\cos\theta_{1}\hat{  z}+\sin\theta_{1}\hat{  x}]~, \nonumber \\
&{\vec q}_{3\omega_0,2}=   \frac{3\omega_0}{c} n_{2}(3\omega_0) [-\cos\theta_{2}\hat{  z}+\sin\theta_{2}\hat{  x}]~.
\end{align}
According to the boundary condition relations of Eq.\ref{eq:BC}, we find $q_{3\omega_0,1,x}=q_{3\omega_0,2,x}=3q_x$. Therefore, we derive the Snell's law for THG:
\begin{align}\label{eq:snell_like}
n_{2}(3\omega_0)\sin\theta_{2} =n_{1}(3\omega_0)\sin\theta_{1}= n_{1}(\omega_0)\sin\theta~.
\end{align}
Considering frequency dependence of the refractive indexes, the Snell's law for THG implies that $\sin\theta_1=[ n_1(\omega_0)/n_1(3\omega_0)] \sin\theta$ is not generally equal to $\sin\theta$. This is in contrast with the specular reflection for first harmonic generation\cite{Jackson_book}.

The plane wave nature of the TH radiations implies:
\begin{align}\label{eq:plane_wave}
\cos\theta_{1} {\cal E}^{(3)}_{1,z}+\sin\theta_{1} {\cal E}^{(3)}_{1,x}=0~,\nonumber \\
-\cos\theta_{2} {\cal E}^{(3)}_{2,z}+\sin\theta_{2} {\cal E}^{(3)}_{2,x}=0~.
\end{align}
By considering Eqs.\ref{eq:p3},\ref{eq:q3tb}, the boundary condition relations Eqs.\ref{eq:BC} become:
\begin{align}
\label{eq:bc_1} &
n_{1}(3\omega_0) \left [\cos\theta_{1}   {\cal E}^{(3)}_{1,x} - \sin\theta_{1}  {\cal E}^{(3)}_{1,z} \right ] +
\nonumber\\
 &n_{2}(3\omega_0) \left [\cos\theta_{2}    {\cal E}^{(3)}_{2,x} +\sin\theta_{2}    {\cal E}^{(3)}_{2,z} \right ]
=i \frac{3\omega_0}{c} \frac{\widetilde{\cal P}_x}{\epsilon_0}~,  \\
\label{eq:bc_2}&
n_{1}(3\omega_0) \cos\theta_{1}   {\cal E}^{(3)}_{1,y}
+
n_{2}(3\omega_0) \cos\theta_{2}    {\cal E}^{(3)}_{2,y}
= -i  \frac{3\omega_0}{c} \frac{\widetilde{\cal P}_y}{\epsilon_0}~,  \\
\label{eq:bc_3}&
n_{1}(3\omega_0) \sin\theta_{1}  {\cal E}^{(3)}_{1,y}
-
n_{2}(3\omega_0) \sin\theta_{2}  {\cal E}^{(3)}_{2,y}  = 0~,  \\
\label{eq:bc_4}&{\cal E}^{(3)}_{1,x} = {\cal E}^{(3)}_{2,x}~,  \\
\label{eq:bc_5}&{\cal E}^{(3)}_{2,y} = {\cal E}^{(3)}_{2,y} ~, \\
\label{eq:bc_6}&
n_{1}(3\omega_0)^2{\cal E}^{(3)}_{1,z} - n_{2} (3\omega_0)^2{\cal E}^{(3)}_{2,z}= -
i \frac{3\omega_0}{c} \frac{\widetilde{\cal P}_x}{\epsilon_0} n_{1}(\omega_0)\sin\theta~.
\end{align}
From Eqs.\ref{eq:bc_1}-\ref{eq:bc_6},\ref{eq:snell_like},\ref{eq:plane_wave} we get:
\begin{align}
 &{\cal E}^{(3)}_{i,x}=  S_{i,x}
\frac{\sigma^{(3)}_{xxxx} }{c\epsilon_0}    \left \{ {\cal E}^3_{x} +{\cal E}_{x} {\cal E}^2_{y}\right \} ~,\\
 &{\cal E}^{(3)}_{i,y}= S_{i,y}
\frac{\sigma^{(3)}_{xxxx} }{c\epsilon_0}    \left \{ {\cal E}^3_{y} +{\cal E}_{y} {\cal E}^2_{x}\right \} ~,\\
 &{\cal E}^{(3)}_{i,z}= S_{i,z}
\frac{\sigma^{(3)}_{xxxx} }{c\epsilon_0}    \left \{ {\cal E}^3_{x} +{\cal E}_{x} {\cal E}^2_{y}\right \} ~.
\end{align}
where
\begin{align}
&S_{1,x}=S_{2,x}=- \frac{\cos\theta_{1}\cos\theta_{2}}{n_{1}(3\omega_0)\cos\theta_{2}+n_{2}(3\omega_0)\cos\theta_{1}} ~,\\
&S_{1,y}=S_{2,y}=-\frac{1}{n_{1}(3\omega_0)\cos\theta_{2}+n_{2}(3\omega_0)\cos\theta_{1}} ~,\\
&S_{1,z}= \frac{\cos\theta_{2}\sin\theta_{1}}{n_{1}(3\omega_0)\cos\theta_{2}+n_{2}(3\omega_0)\cos\theta_{1}} ~,\\
&S_{2,z}=  -\frac{\cos\theta_{1}\sin\theta_{2}}{n_{1}(3\omega_0)\cos\theta_{2}+n_{2}(3\omega_0)\cos\theta_{1}}~.
\end{align}
For normal incidence we have $\theta=0$. From Eq.\ref{eq:snell_like} we have $\theta_1=\theta_2=0$. Therefore, $S_{i,z}=0$ and $S_{i,x}=S_{i,y}=-1/[n_{1}(3\omega_0) +n_{2}(3\omega_0)]$. The time-average of the incident intensity gives $I_{\omega_0} = 2 n_1(\omega_0) \epsilon_0 c |{\vec E}_{\rm in}|^2 $. The intensity of the transmitted TH signal is $I_{3\omega_0} = 2 n_2(3\omega_0) \epsilon_0 c |{\vec E}^{(3)}|^2$. From this we get Eq.\ref{eq:THG_efficiency} for THGE.
\subsubsection{Symmetry considerations}
\label{subss:symmetry}
The rank-4 tensor of $\sigma^{(3)}$ transforms as follows under an arbitrary $\phi$-rotation:
\begin{align}\label{eq:sym}
\sigma^{(3)}_{\alpha'\beta'\gamma'\delta'}=\sum_{\alpha\beta\gamma} R_{\alpha'\alpha}(\phi) R_{\beta'\beta}(\phi) R_{\gamma'\gamma}(\phi) R_{\delta'\delta}(\phi) \sigma^{(3)}_{\alpha\beta\gamma\delta}~.
\end{align}
We take the $z$-axis as the rotation-axis, perpendicular SLG. Therefore, the rotation tensor is:
\begin{equation}\label{eq:R}
\overset\leftrightarrow R(\phi) = \begin{pmatrix} \cos\phi &\sin\phi \\ -\sin\phi &\cos\phi \end{pmatrix}~.
\end{equation}
We take $\hat{\bm \ell}= \overset\leftrightarrow R(\phi)\cdot\hat{\bm x}$. By plugging Eq.\ref{eq:R} in \ref{eq:sym}, we get:
\begin{align}\label{eq:rot}
\sigma^{(3)}_{\ell\ell\ell\ell} & =
 [\sin\phi]^4 \sigma^{(3)}_{yyyy} +[\cos\phi]^4 \sigma^{(3)}_{xxxx}
  \nonumber\\&
+\cos\phi [\sin\phi]^3 \left [ \sigma^{(3)}_{xyyy} +\sigma^{(3)}_{yxyy} +\sigma^{(3)}_{yyxy} +\sigma^{(3)}_{yyyx} \right ]
 \nonumber\\&
+[\cos\phi]^3\sin\phi \left [\sigma^{(3)}_{xxxy}+\sigma^{(3)}_{xxyx}+\sigma^{(3)}_{xyxx}+\sigma^{(3)}_{yxxx} \right ]
\nonumber\\&
+ [\cos\phi \sin\phi]^2 \big  [\sigma^{(3)}_{xxyy} +\sigma^{(3)}_{xyxy}+\sigma^{(3)}_{xyyx}
\nonumber\\&
+ \sigma^{(3)}_{yxxy}+ \sigma^{(3)}_{yxyx} +\sigma^{(3)}_{yyxx}  \big].
\end{align}
Because of the $C_{6v}$ symmetry for SLG on a substrate, there are only 4 independent tensor elements\cite{Boyd2003}:
\begin{align}
&\sigma^{(3)}_{xxxx}=\sigma^{(3)}_{yyyy}=\sigma^{(3)}_{xxyy}+\sigma^{(3)}_{xyyx}+\sigma^{(3)}_{xyxy}
\nonumber\\&
\sigma^{(3)}_{xxyy}=\sigma^{(3)}_{yyxx},
\nonumber\\&
\sigma^{(3)}_{xyyx}=\sigma^{(3)}_{yxxy},
\nonumber\\&
\sigma^{(3)}_{xyxy}=\sigma^{(3)}_{yxyx}.
\end{align}
By implementing Eq.43 in Eq.\ref{eq:rot}, we get $\sigma^{(3)}_{\ell\ell\ell\ell}=\sigma^{(3)}_{xxxx}$.
\subsubsection{Effect of finite relaxation rate}
\label{subss:Gamma}
\begin{figure}
\centerline{\includegraphics[width=80mm]{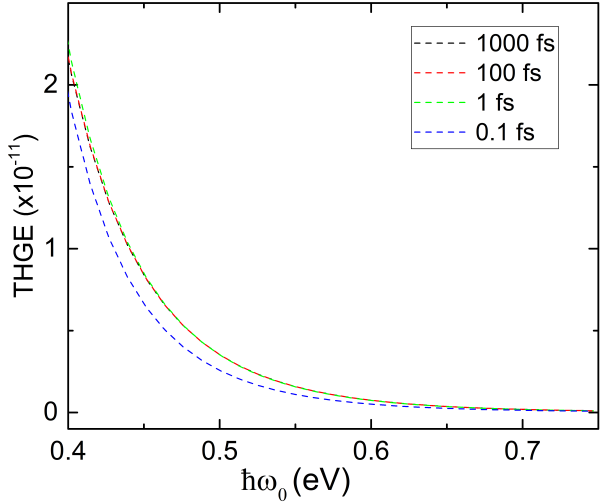}}
\caption{THGE for SLG on Sa as a function of $\omega_0$ for different $\tau=\hbar/\Gamma$ at constant $T_e$=2000K and $E_F$=200meV, for incident intensity$\sim2.4\times 10^{12}{\rm Wm}^{-2}$, corresponding to the value used in our experiments}
\label{fig:THGE_tau}
\end{figure}

The effect of finite $\tau$ in the TH conductivity can be derived from\cite{ChengPRB2015}:
\begin{widetext}
\begin{eqnarray}\label{eq:clean}
\bar{\sigma}^{(3)}_{xxxx} (\omega_0,E_{\rm F},0) & \approx &
\frac{ 17  G(2|E_{\rm F}|,\hbar\omega_0+i\Gamma)  - 64 G(2|E_{\rm F}|,2\hbar\omega_0+i\Gamma) + 45 G(2|E_{\rm F}|,3\hbar\omega_0+i\Gamma)}{24 (\hbar\omega_0)^4}
\nonumber \\
&+&\frac{\Gamma}{6(\hbar\omega_0)^4}
 \Bigg \{
 17 \left [\frac{1}{2  | E_{\rm F} | +3 \hbar\omega_0 +i \Gamma }+\frac{1}{2  | E_{\rm F} | -3 \hbar\omega_0 -i \Gamma}\right ]
  \nonumber \\
 &-& 8 \left [ \frac{1}{2 | E_{\rm F} |  +2 \hbar\omega_0 +i \Gamma}+\frac{1}{2  | E_{\rm F} | -2 \hbar\omega_0 -i \Gamma}\right ]
 \nonumber\\
 &+& 3 \hbar\omega_0  \left [\frac{1}{(2  | E_{\rm F} | +3 \hbar\omega_0 +i \Gamma )^2}-\frac{1}{(2 | E_{\rm F} | -3 \hbar\omega_0 -i \Gamma  )^2}\right ]
\Bigg \}~.
\end{eqnarray}
\end{widetext}
Note that ($\approx$) is because we assume $\Gamma\ll \hbar\omega_0$\cite{ChengPRB2015}. Fig.\ref{fig:THGE_tau} shows that a finite $\tau$ has a small effect on THGE for most of the SLG in literature, as well as those in this paper.
\subsubsection{$T_{\rm e}$ and $E_{\rm F}$ effects on THGE}
\label{subss:THGE_TeandEf}
\begin{figure}
\centerline{\includegraphics[width=90mm]{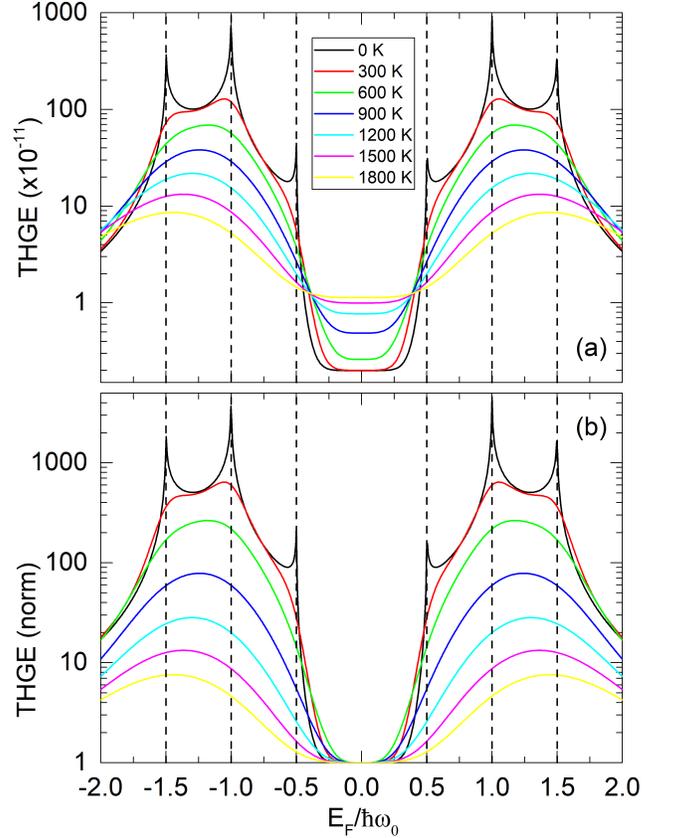}}
\caption{\label{fig:THGE_Te} $E_{\rm F}$ dependence of THGE for SLG on SiO$_2$ at $\hbar\omega_0=500{\rm meV}$ for different $T_{\rm e}$ between 0K and 1800K. (a) Absolute THGE. (b) THGE normalized to the minimum so that THGE at $E_{\rm F} = 0$ is equal to 1 for all $T_{\rm e}$.}
\end{figure}
\begin{figure}
\centerline{\includegraphics[width=80mm]{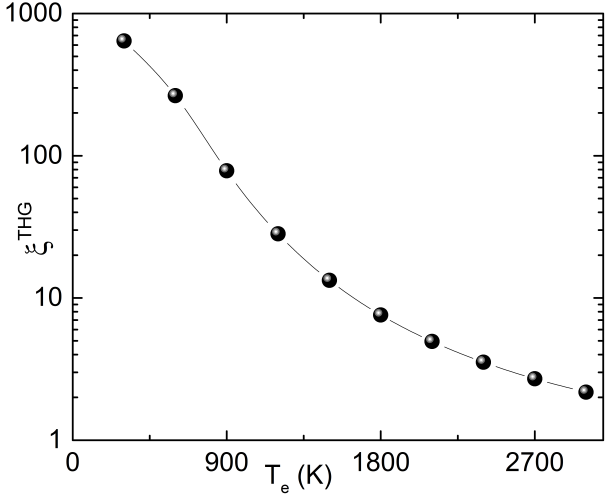}}
\caption{\label{fig:THGE_xi} $T_e$ dependence of $\xi^{\rm THG}$.}
\end{figure}
The $T_e$ and $E_F$ dependence of THGE for SLG on SiO$_2$ at $\hbar\omega_0=500$meV is shown in Fig.\ref{fig:THGE_Te}, where 3 logarithmic singularities at $2|E_F|=\hbar\omega_0,2\hbar\omega_0, 3\hbar\omega_0$ for $T_e$=0K can be seen. By increasing $T_e$, the first peak at $2|E_F|=\hbar\omega_0$ disappears and the two others merge and form a broad maximum, roughly located at $2|E_{\rm F}|\sim (2+3)\hbar\omega_0/2=2.5\hbar\omega_0$. THGE is almost insensitive to $E_{\rm F}$ for $2|E_{\rm F}|<\hbar\omega_0$. This can be explained using the asymptotic relation of the TH conductivity for $|E_{\rm F}|\ll \hbar\omega_0 $. For $T_{\rm e}=0$:
\begin{align}
\label{eq:sigma3xxxx}
\sigma^{(3)}_{xxxx} \approx
\frac{e^4 \hbar v^2_{\rm F}}{(\hbar\omega_0)^4} \left \{ \frac{1}{96} +
\frac{i}{\pi}  \left(\frac{2|E_{\rm F}|}{3\hbar\omega_0} \right)^3+\dots \right \}
\end{align}
Eqs.\ref{eq:sigma3xxxx},\ref{eq:THG_efficiency} explain the flat part of the curves in Fig.\ref{fig:THGE_Te} in the low-doping regime ($\hbar\omega_0>$2$|E_F|$).

In order to quantify the tuneability of THG in SLG by altering $E_F$, we define a parameter:
\begin{align}
\xi^{\rm THG} \equiv  \frac{\eta^{\rm THG}_{\rm max} }{\eta^{\rm THG}_{\rm min}}~,
\end{align}
where $\eta^{THG}_{min}$ stands for THGE in the nearly undoped regime ($|E_F|\ll \hbar\omega_0$). Fig.\ref{fig:THGE_xi} indicates that $\xi^{THG}$ decreases by increasing $T_e$.
\subsection{Fermi energy, Fermi level, chemical potential and the estimation of $T_e$ in photoexcited SLG}
\label{subs:Te}
\begin{figure}
\centerline{\includegraphics[width=90mm]{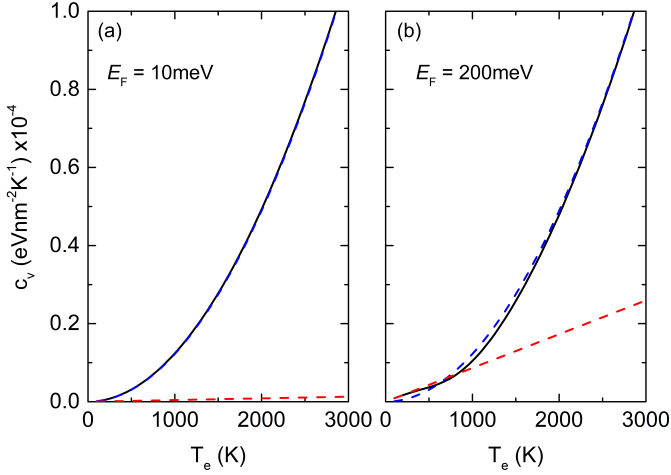}}
\caption{\label{fig:cv}
Numerical calculation of $c_v$ for (a) $E_F$=10 and (b) 300meV. The blue and red dashed lines are Eqs.\ref{eq:cv_undoped}, \ref{eq:cv_doped}.}
\end{figure}
When a pulsed laser interacts with SLG, for several hundred fs after the pump pulse, the electron and hole distributions in valence and conduction bands are given by the Fermi-Dirac functions $f_{\rm FD}(\varepsilon; \mu_{\lambda}, T_{\rm e})$ with the same $T_{\rm e}$ and two chemical potentials $\mu_{\rm v}$ and $\mu_{\rm c}$. By definition $\mu_{\rm c}=-\mu_{\rm v}$. The term Fermi level ($E_{\rm FL}$) is sometimes used in literature to denote $\mu$. The Fermi energy ($E_{\rm F}$) is defined as the value of $\mu$ at $T_{\rm e}=0K$. $E_{\rm F}$ is thus a function of the electron density only. Only after the recombination of the photoexcited electron-hole pairs, a single Fermi-Dirac distribution is established in both bands and the equilibrium condition $\mu_{\rm v} =\mu_{\rm c}$ holds. The recombination time depends on carrier density and laser fluence, and can be much longer than the time$\lesssim 20fs$ needed for thermalization (see Ref.\citenum{tomadin_prb_2013} and references therein).

\begin{figure}
\centerline{\includegraphics[width=90mm]{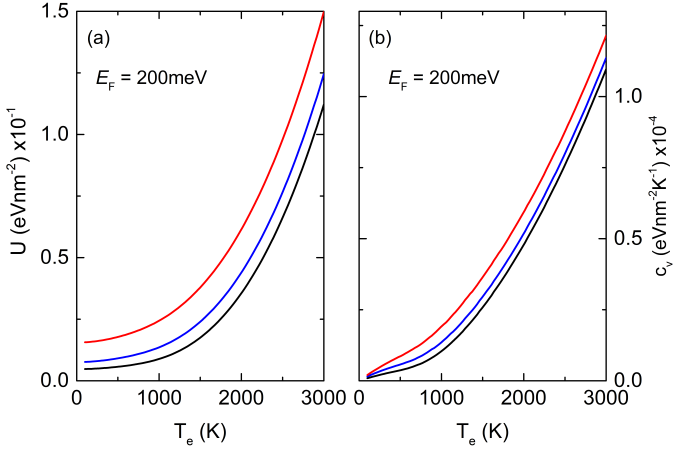}}
\caption{\label{fig:cv_pe}
Numerical calculation of (a) electron energy density and (b) $c_v$ for $E_F$=200meV. The blue, and red lines correspond to photoexcited densities $\delta n_e= 10^{12}~{cm}^{-2}$ and $\delta n_e=3 \times 10^{12}~{cm}^{-2}$, while the black line corresponds to a thermalized system with a single $\mu$}
\end{figure}

$N_{\rm f}=4$ is the product of spin and valley degeneracy; $\nu(\varepsilon)=N_{\rm f}|\varepsilon|/[2 \pi (\hbar v_{\rm F})^{2}]$ the density of electronic states per unit of area. The electronic heat capacity $c_{\rm v}$ is defined as the derivative of the electronic energy density $U$ with respect to $T_{\rm e}$. This quantity depends on all the variables which affect the electronic energy density, such as $T_{\rm e}$ and the carrier density, or equivalently $\mu_{\rm c}$ and $\mu_{\rm v}$. In a photoexcited system, in general, $c_{\rm v}$ depends on both the electron and hole densities, \emph{i.e.} on both $\mu_{\rm c}$ and $\mu_{\rm v}$. In this case, $c_{\rm v}$ can be written as\cite{Grosso_book}:
\begin{align}\label{eq:cv_full}
c_{\rm v}(\mu_{\rm c}, \mu_{\rm v}, T_{\rm e})
&= \frac{\partial}{\partial T_{\rm e}} \int_{0}^{\infty} d\varepsilon \nu(\varepsilon) \varepsilon f_{\rm FD}(\varepsilon; \mu_{\rm c}, T_{\rm e})
\nonumber\\
&+ \frac{\partial}{\partial T_{\rm e}} \int_{0}^{\infty} d\varepsilon \nu(\varepsilon) \varepsilon f_{\rm FD}(\varepsilon; -\mu_{\rm v}, T_{\rm e})~,
\end{align}
where the first integral is the electron and the second the hole contribution, and the Fermi-Dirac distribution is
\begin{equation}
f_{\rm FD}(\varepsilon; \mu, T_{\rm e}) = \frac{1}{e^{(\varepsilon-\mu) / (k_{\rm B} T_{\rm e})} + 1}~.
\label{eq:fFD}
\end{equation}
To take the $T_{\rm e}$ derivative in Eq.\ref{eq:cv_full}, the $T_{\rm e}$ dependence of $c_{\rm v}$ has to be specified. The electron and hole densities are given by:
\begin{align}
&n_{\rm e}(\mu_{\rm c}, T_{\rm e}) = \int_{0}^{\infty} d\varepsilon \nu(\varepsilon) f_{\rm FD}(\varepsilon; \mu_{\rm c}, T_{\rm e}), \quad
\nonumber\\
&n_{\rm h}(-\mu_{\rm v}, T_{\rm e}) = \int_{0}^{\infty} d\varepsilon \nu(\varepsilon) f_{\rm FD}(\varepsilon; -\mu_{\rm v}, T_{\rm e})~.
\label{eq:nEnH}
\end{align}
Since the total electron density in both bands is constant, the difference between electron and hole densities is constant:
\begin{equation}
\label{eq:cons_electr}
n_{\rm e}^{(0)} - n_{\rm h}^{(0)} = n_{\rm e}(\mu_{\rm c}, T_{\rm e}) - n_{\rm h}(-\mu_{\rm v}, T_{\rm e})~,
\end{equation}
where $n_{\rm e}^{(0)}$ and $n_{\rm h}^{(0)}$ are the electron and hole densities. At equilibrium, when $\mu_{\rm c}=\mu_{\rm v}=\mu$, Eq.\ref{eq:cons_electr} can be solved for $\mu$. A photoexcited density $\delta n_{\rm e}$ changes the densities in both bands as follows:
\begin{align}\label{eq:pe_dens}
&n_{\rm e}(\mu_{\rm c}, T_{\rm e}) = n_{\rm e}(\mu, T_{\rm e}) + \delta n_{\rm e}, \quad
\nonumber\\
&n_{\rm h}(-\mu_{\rm v}, T_{\rm e}) = n_{\rm h}(-\mu, T_{\rm e}) + \delta n_{\rm e}~.
\end{align}
After finding $\mu$ with Eq.\ref{eq:cons_electr}, one can get $\mu_{\rm c}$ and $\mu_{\rm v}$ with Eq.\ref{eq:pe_dens}. This defines the $c_{\rm v}$ dependence of $T_{\rm e}$ in Eq.\ref{eq:cv_full}, and allows to calculate the temperature derivative. The result of Eq.\ref{eq:cv_full} is shown in Fig.\ref{fig:cv} for $\mu_{\rm c}=\mu_{\rm v}=\mu$. In Ref.\citenum{LuiPRL2010} the following expression is given for $c_{\rm v}$:
\begin{equation}\label{eq:cv_undoped}
c_{\rm v}(T_{\rm e}) = \frac{18 \zeta(3)}{\pi (\hbar v_{\rm F})^{2}} k_{\rm B}^{3} T_{\rm e}^{2}~.
\end{equation}
In principle, as noted in Ref.\citenum{SunNN2012}, Eq.\ref{eq:cv_undoped} is valid at the charge neutrality point $|\mu| \ll k_{\rm B} T$ only. For a degenerate system, $k_{\rm B} T \ll |\mu|$, we have\cite{Vignale_book}:
\begin{equation}\label{eq:cv_doped}
c_{\rm v}(\mu, T_{\rm e}) = \frac{\pi^{2}}{3} \nu(E_{\rm F}) k_{\rm B}^{2} T_{\rm e}~,
\end{equation}
as derived \emph{e.g.} in Eqs.8.10 of Ref.\citenum{Vignale_book}, in Eq.4 of Ref.\citenum{LowPRB2012} and in Eq.8 in the Supplementary Information of Ref.\citenum{TielNP2013}. However, the numerical calculation in Fig.\ref{fig:cv} shows that the quadratic approximation (Eq.\ref{eq:cv_undoped}) is much better in the regime where $T_{\rm e}\sim 1000K$ and $\mu \sim 100$meV. Fig.\ref{fig:cv_pe} shows that, taking into consideration the difference between $\mu_c$ and $\mu_v$, for typical values of the photoexcited density, contributes$\gtrsim 15\%$ to $c_v$.

The number of photoexcited electron-hole pairs per unit area in the time interval $dt$ is given by the number of absorbed photons in the same time interval per unit area, i.e. $(d n_{\rm e} + d n_{\rm h})/2 = (P / A) / (\hbar \omega_0) d t $. In the steady state, the energy delivered by the pump is transferred into the phonon modes. Hence, we identify the electron-hole recombination time with $\tau$. We then get:
\begin{align}
&\frac{1}{2} \left ( \frac{d n_{\rm e}}{dt} + \frac{d n_{\rm h}}{dt} \right ) = \frac{1}{\hbar \omega_0} \frac{P}{A}
\nonumber\\
&- \frac{1}{2}\frac{[n_{\rm e}(\mu_{\rm c}, T_{\rm e}) + n_{\rm h}(-\mu_{\rm v}, T_{\rm e})] - (n_{\rm e}^{(0)} + n_{\rm h}^{(0)})}{\tau}~.
\end{align}
In the steady state this becomes:
\begin{equation}\label{eq:e_h_pairs}
n_{\rm e}^{(0)} + n_{\rm h}^{(0)} = n_{\rm e}(\mu_{\rm c}, T_{\rm e}) + n_{\rm h}(-\mu_{\rm v}, T_{\rm e}) - \frac{2 \tau}{\hbar \omega_0} \frac{P}{A}.
\end{equation}
Combining Eqs.\ref{eq:cons_electr},\ref{eq:e_h_pairs}, we find:
\begin{equation}
\delta n_{\rm e} = \frac{\tau}{\hbar \omega_0} \frac{P}{A}~.
\end{equation}
To calculate $E_F$ (e.g. for a $n$-doped sample) one needs to solve Eqs.\ref{eq:fFD}, \ref{eq:nEnH}, \ref{eq:cons_electr} with $\mu_{\rm c}=\mu_{\rm v}=E_F$, $T_e=0$, and $n_{h}^{(0)}=0$, finding $E_F=\hbar v_{\rm F} \sqrt{\pi n_{\rm e}}$. This relation can be safely used at $T_{\rm e}=300K$ and electron densities $n_{\rm e}^{(0)} \gtrsim 10^{11}$ because the density of thermally excited holes is negligible. Indeed, experimental measurements of the carrier density of $n$-doped SLG at room temperature routinely neglect the hole population contributions. In a photoexcited SLG, even after the recombination of the photoexcited electron-hole pairs, the $T_e$ dependence of $\mu$ cannot be ignored. In this case, to calculate $\mu$, one needs to solve Eqs.\ref{eq:fFD},\ref{eq:nEnH}, \ref{eq:cons_electr} with $\mu_{\rm c}= \mu_{\rm v}=\mu$ and $n_{\rm h}^{(0)}=0$ as a function of $T_{\rm e}$. This gives $\mu=E_F [1 - \pi^{2} T_{\rm e}^{2} / (6 T_{\rm F}^{2})]$ for $T_{\rm e}\lesssim T_{\rm F}$ and $\mu =E_F T_{\rm F}/ (4 {\rm ln}2 \times T_{\rm e})$ for $T_{\rm e}\gtrsim T_{\rm F}$\cite{hwang_prb_2009}, where $T_{\rm F}=E_{\rm F}/K_{\rm B}$, with K$_{\rm B}$ the Boltzmann constant. For a typical case of $E_F$=200meV and $T_e=$1500K, we have $\mu \sim 0.3-0.5 E_F$.

\section*{\label{Ackn}Acknowledgements}
We acknowledge funding from EU Graphene Flagship, ERC Grant Hetero2D, EPSRC Grants EP/K01711X/1, EP/K017144/1, EP/N010345/1, and EP/L016087/1.


\begin{thebibliography}{100}
\bibitem{ref:proof}
By definition, $\mu_e=\sigma/(n e)$ where $n$ is the carrier density and $\sigma=D/(\pi\Gamma)$ is the Drude conductivity of graphene in which $D=e^2E_{\rm F}/\hbar$\cite{Abedinpour_prb_2011} is the Drude weight. From Eq.5 we get  $n\pi=[E_{\rm F}/(\hbar v_{\rm F})]^2$, $\Gamma=e\hbar v^2_{\rm F}/(\mu_e E_{\rm F})$. Note that this relation is not valid for the undoped limit when $E_{\rm F}/\Gamma \to  0$, where inter-band effects are dominant and the Drude modeling is not sufficient.
\end{thebibliography}
\end{document}